\documentclass{nsart_eng}
\usepackage{cite}
\usepackage[dvips]{graphicx}
\usepackage{amsmath,amssymb}

\year{2017} \volume{8} \nomer{1} \firstpage{1}

\let\le\leqslant

\let\ge\geqslant

\begin{document}

\setcounter{page}{1}
\newcommand{\re}[1]{(\ref{#1})}
\newcommand{\lab}[1]{\label{#1}}
\newcommand{\ci}[1]{\cite{#1}}
\renewcommand{\baselinestretch}{1.25}
\newcommand{\bfr}{\begin{flushright}}
\newcommand{\bfl}{\begin{flushleft}}
\newcommand{\efl}{\end{flushleft}}
\newcommand{\efr}{\end{flushright}}
\newcommand{\bc}{\begin{center}}
\newcommand{\ec}{\end{center}}
\newcommand{\be}{\begin{equation}}
\newcommand{\ee}{\end{equation}}
\newcommand{\bea}{\begin{eqnarray}}
\newcommand{\eea}{\end{eqnarray}}
\newcommand{\ba}{\begin{array}}
\newcommand{\ea}{\end{array}}
\newcommand{\edc}{\end{document}}
\newcommand{\ul}{\underline}
\newcommand{\ri}{\rightarrow\infty}
\newcommand{\li}{\leftarrow\infty}
\newcommand{\ra}{\rightarrow}
\newcommand{\la}{\leftarrow}
\newcommand{\ds}{\displaystyle}
\newcommand{\dsf}{\displaystyle\frac}
\newcommand{\dt}{\Delta{t}}
\newcommand{\il}{\int\limits}
\newcommand{\pal}{\partial}
\newcommand{\cal}{\mathcal}
\newcommand{\bone}{{\bf 1}}
\newcommand{\gComment}[1]{}
\renewcommand{\gComment}[1]{\textcolor{red}{Gerardo: #1}}


\title[{\it Metal-insulator (fermion-boson)-crossover origin...}]
{Metal-insulator (fermion-boson)-crossover origin of pseudogap phase of 
cuprates I: anomalous heat conductivity, insulator resistivity boundary, 
nonlinear entropy}

\author[{\it B.~Abdullaev, C.\,-H.~Park, K.\,-S.~Park, I.\,-J. Kang}]
{B.~Abdullaev$^1$, C.\,-H.~Park$^2$, K.\,-S.~Park$^3$, I.\,-J. Kang$^4$}

\address{
$^1$Institute of Applied Physics, National University of
Uzbekistan, Tashkent 100174, Uzbekistan.\\
$^2$Research Center for Dielectric and Advanced Matter Physics,
Department of Physics, Pusan National University, 30
Jangjeon-dong, Geumjeong-gu, Busan 609-735, Korea.\\
$^3$Department of Physics, Sungkyunkwan University, 2066,
Seoburo, Jangangu, Suwon, Gyeonggido, Korea.\\
$^4$Samsung Mobile Display, Suwon, Kyunggido, Korea.}

\email{bakhodir.abdullaeff@yandex.ru, cpark@pusan.ac.kr, pmpark@skku.edu, hacret@hanmailnet}







\begin{abstract}
Among all experimental observations of cuprate physics,
the metal-insulator-crossover (MIC), seen in the pseudogap (PG) region of the 
temperature-doping phase diagram of copper-oxides under a strong magnetic field, 
when the superconductivity is suppressed, is most likely the most intriguing one. 
Since it was expected that the PG-normal state for these materials, as for conventional 
superconductors, is conducting. This MIC, revealed in such phenomena as heat conductivity 
downturn, anomalous Lorentz ratio, insulator resistivity boundary, nonlinear entropy, 
resistivity temperature upturn, insulating ground state, nematicity- and stripe-phases 
and Fermi pockets, unambiguously indicates on the insulating normal state, from which the 
high-temperature superconductivity (HTS) appears. In the present work (article I), we discuss the MIC phenomena 
mentioned in the title of article. The second work (article II) will be devoted to discussion 
of other listed above MIC phenomena and also to interpretation of the recent observations
in the hidden magnetic order and scanning tunneling microscopy (STM) experiments 
spin and charge fluctuations as the intra PG and HTS pair ones. We find that all these 
MIC (called in the literature as non-Fermi liquid) phenomena can be obtained within 
the Coulomb single boson and single fermion two liquid model, which we recently developed, 
and the MIC is a crossover of single fermions into those of single bosons. We show that
this MIC originates from bosons of Coulomb two liquid model and fermions, whose origin is
these bosons. At an increase of doping up to critical value or temperature up to PG boundary 
temperature, the boson system undegoes bosonic insulator - bosonic metal - fermionic 
metal transitions. 
\end{abstract}



\maketitle


\section{Introduction}
\label{sec1}

The origin of PG and HTS phases in copper oxides (cuprates) is one of the most puzzling 
and challenging problem in condensed matter physics. Despite being almost three decades 
since their discovery, intensive experimental and theoretical studies have yielded little 
clear understanding of these phases so far.
The experimental studies of HTS and PG in cuprates have provided physicist 
by numerous interesting and fascinating materials with unconventional properties. Among 
the most puzzling and thus far most intriguing is the observation of the 
MIC, seen in the underdome region of a temperature-doping phase diagram in the 
absence~\ci{Takagi,Keimer,Wuyts,Abe} or presence~\ci{Ando,Boebinger,Fournier,Ono,Ando1} of a strong external 
magnetic field. The MIC, detected after suppression of the HTS by a strong magnetic field, 
results in a number of different phenomena: heat conductivity downturn and anomalous Lorentz 
ratio~\ci{hill,proust}, insulator resistivity boundary~\ci{Boebinger}, 
nonlinear entropy~\ci{loram1,loram2}, insulating ground state (see Refs~\ci{Takagi,Keimer,Wuyts,Abe} and 
~\ci{Ando,Boebinger,Fournier,Ono,Ando1}), dynamic nematicity~\ci{Fujita} and 
static stripe phases~\ci{Vojta,Vojta1} and Fermi pockets~\ci{Vojta,Vojta1,Sebastian}. 
This reveals the highly unconventional dielectric properties of the PG-normal phase of these 
superconductors. Since superconductivity appears in conventional superconductors from 
the conducting normal state only, the understanding of how HTS arises from an insulating state 
becomes a fundamental problem, and thus, the keystone for cuprate physics. This 
MIC also separates previously applied mechanisms and models for conventional superconductors 
from the consideration. 

Being previously introduced for the description of an electron transport transition of metals 
into insulators, as we see below in this and also in the second II papers, a concept of the MIC
provides an apportunity to encompass various properties of the famous cuprate PG physics in the 
understanding withing a single treatment. his treatment might be
our Coulomb single boson and single fermion two liquid model, which was developed in Ref.~\ci{Abdullaev1}.

Analyzing the electric charge and percolation of the visualized in STM real space PG and HTS 
nanoregions (NRs) of the $Bi_2Sr_2CaCu_2O_{8+\delta}$ compound~\ci{Gomes,Pan}, which exhibited 
an energy gap, we showed in~\ci{Abdullaev1} that NRs with minimal size are PG and HTS pairs 
and furthermore, they are single bosons, and percolation of these NRs occurs from the first 
critical doping, from which the HTS begins. In~\ci{Abdullaev1} we were also able to qualitatively 
understand all elements for the temperature-doping phase diagram of the copper-oxide. If the
experiment~\ci{Gomes} clearly showed, through the visualization of pair formation in the PG
region, the precursor mechanism of the bulk HTS, we have successfully realized that
the later appears due to percolative overlapping of the separated NRs. All positions of the
Coulomb single boson and single fermion two liquid model are listed in Ref.~\ci{Abdullaev7}.
This model is semi-phenomenological, since if the microscopic formation of single bosons, through the anyon
bosonization of 2D fermions, is rigorously proven and after that experimentally confirmed in~\ci{Abdullaev1}, 
then the size of these bosons and percolative nature of the HTS are the results of Refs.~\ci{Gomes,Pan} data analysis.   

In the present paper, we describe the low-temperature (low-$T$) non-Fermi liquid
behavior of the heat-transport and entropy. As will be seen below, the central role 
in this calculation plays the temperature $T$ dependence of the electronic specific heat $c$.
This $T$ dependence of $c$ is a result of the insulating ground state of 2D gas of plasmons, 
which consisted of charged single bosons, deformed and pinned by cuprate parent compound lattice ferroelectric  
atoms (the nematicity electronic state (see for details Ref.~\ci{Abdullaev7} and paper II)).
Since, according to the Bogoliubov approach for gas of charged bosons~\ci{Abdullaev5}, 
at high gas density, where this approach is valid, the ground state energy consists of 
components for the Bose-Einstein condensate and a gas of quasi-particle-plasmons. 
Following March {\it et} {\it al.}~\ci{march}, who showed that the Bogoliubov approach can be nicely 
applied to the description of the liquid $He$ II, we believe that this approach
may at least qualitatively desribe the intermediate densities of 2D Coulomb boson gas, at which
the HTS takes place. At high magnetic fields or at lower dopings, close to the first 
critical doping, for which MIC phenomena are measured, the Bose-Einstein condensate vanishes 
and there only a gas of quasi-particle-plasmons exists. 
However, the latter, the individual charges of which  being fixed by ferroelectric atoms, 
is insulating therefore, the insulator is the whole ground state of copper oxides. Typically, 
the experiment detects the MIC up to the critical value of 
doping~\ci{Boebinger,Fournier,Ono}, which, for some cuprates, is close to the optimal 
doping, where the $T_c$ is maximum in the temperature-doping phase diagrams. There also exists 
the case of $Bi_2Sr_2CaCu_2O_{8+\delta}$ compound~\ci{Gomes}, for which this critical doping 
almost coincides with the second critical doping, at which the HTS vanishes.

We start with description of existing experimental data for the
heat-transport and PG-normal state specific heat of copper-oxides.
Hill with collaborators \ci{hill} reported that the heat
conductivity of the electron doped copper-oxide $Pr_{2-x} Ce_x
CuO_4$ measured at low-T deviated from one predicted by the
Landau Fermi liquid theory (LFLT), i.e., as the temperature decreases, the temperature
dependence of the heat conductivity ($\kappa$) is changed from the
normal linear $\kappa \sim T$ behavior into an anomalous $T^{3.6}$
one, which was described by the "downturn" behavior of the heat
conductivity. They also reported another important non-Fermi liquid behavior:
the Lorentz ratio of the
Wiedemann -- Franz law (WFL) in the region of the linear
T-dependence of $\kappa$ was significantly larger (1.7 times) than
Sommerfeld's value. These violations were also observed in the
heavily over-doped non-superconducting compound $La_{2-x} Sr_x
CuO_4$ by Nakamae $et$ $al.$ \ci{nakamae} and  in $Bi_{2+x}
Sr_{2-x} CuO_{6+\delta}$ copper-oxide in the vicinity of the MIC
by Proust $et$ $al.$ \ci{proust}. Smith $et$ $al.$ suggested that
the downturn behavior of $\kappa$ results solely from the decoupling
of the heat carrying thermal phonons and electrically conducting
charge carriers~\ci{Smith} at low-T, while Hill $et$ $al.$ have
indicated that the downturn of $\kappa$ should be intrinsic for
copper-oxides (see Ref. \ci{hill}).

The normal state electronic specific heat $c$ of superconductors
$Y Ba_2 Cu_3 O_{6+x}$ and $La_{2-x} Sr_x CuO_4$ above the HTS
transition temperature $T_c$ was experimentally investigated in
\ci{loram1} and \ci{loram2}, respectively. A magnetic field
dependence of $c$ has considered in \ci{luo} for
$Y_{0.8}Ca_{0.2}Ba_2 Cu_3 O_{6+x}$ compound. Due to existence of
HTS, it is impossible to extract the information on the low-$T$
dependence of the normal state $c$. On the other hand, Loram $et$
$al.$ \ci{loram1} showed the $T$-dependence of the entropy (${\cal
S}$) ${\cal S} \sim T^i$ with $i>1$ for the underdoped
(insulating) material, which was driven from the measured
electronic specific heat of HTS superconductors $Y Ba_2 Cu_3
O_{6+x}$, ignoring the superconducting effects, while for the
optimal doping compound ${\cal S} \sim T$ was measured. The simple
Drude model for the heat conductivity $\kappa$ will connect in our
consideration the kinetic heat-transport property of PG region of
HTS superconductors with its thermodynamic quantity -- the
electronic specific heat $c$. The entropy ${\cal S}$ is a result
of $c$.

The low-$T$ MIC was observed firstly by Ando $et$ $al.$~\ci{Ando} in 
the hole doped $La_{2-x} Sr_x CuO_4$ compound through
the measurement of the resistivity in $CuO_2$ plane. This paper
suggested the low-$T$ MIC is an intrinsic
property rather than one induced by the 2$D$ localization of
carriers. The doping versus $T$ diagram of the insulating state up
to near-optimal-doping was obtained in the extensive investigation
by Boebinger $et$ $al.$~\ci{Boebinger}. A similar
(near-optimal-doping) observation of the low-$T$ MIC was
reported for the electron doped  $Pr_{2-x} Ce_x CuO_4$ by Fournier
$et$ $al.$~\ci{Fournier} and for the $1/8$-doped $Bi_2 Sr_{2-x}
La_x CuO_{6+\delta}$ by Ono $et$  $al.$ ~\ci{Ono}. The
low-$T$ MIC were clearly measured  in a $La$-free hole-doped
$Bi_{2+x} Sr_{2-x} Cu_{1+y}O_{6+\delta}$ of the doping  $x=0.13$
(while optimal doping was $x=0.16$) by Vedeneev and Maude
\ci{vedeneev} through the measurement of the in-plane resistivity.
Although in some papers (see, for example, Ref. \ci{vedeneev})
authors tried to invoke the 2$D$ carrier localization approach to
explain the phenomenon, the material-independent and universal
character of the low-$T$ MIC for all cuprates indicates
that effect is intrinsic and may be related to a fundamental
property of cuprates.

It was already mentioned above that
according to Ref. \ci{Smith} the downturn of the heat conductivity is
related to a decoupling of charges and phonons. However, the paper
\ci{Smith} does not clarify the origin of this decoupling.
Observations of the low-$T$ MIC, and the downturn for the
same compound and under same experimental conditions are
not occasional and indicate on the MIC origin of the decoupling, when
charges localized in the insulating state cease to do a scattering
with phonons. Therefore, the non-linear
insulating ${\cal S} \sim T^i$ with $i>1$ dependence of entropy,
the downturn behavior $\kappa \sim T^{3.6}$ and the low-$T$ MIC are 
results of the same underlying physics of the insulating ground state 
and MIC for cuprates. 

In this work, we demonstrate that the unified description of the three MIC
phenomena is possible. In Sec. \ref{sec2} we derive the
specific heat of the ideal gas of Bogoliubov quasi-particles (IGBQ) and 
demonstrate that it is obtained from the
contributions of plasmons at low-$T$ and free quasiparticles at
high- $T$ temperatures. We will find in the section $T$, at which
one part of the specific heat transits into another one. In the
Sec. \ref{sec3} the low-$T$ dependence and the downturn
temperature of the heat conductivity $\kappa$ will be determined.
In this section, we also obtain the Lorentz ratio of the WFL and
discuss the arguments supporting our approach for the explanation
of the observed insulating ground state (IGS) and low-$T$ MIC. Analysis of the experimental
data on the specific heat for copper-oxides and their possible
relation to $c$ described in the present paper will be made in
Sec. \ref{sec4}. In this section we obtain also the expression for
entropy  ${\cal S}$.  We conclude and summarize the paper
by Sec. \ref{sec5}.

\section{Specific heat of IGBQ}
\label{sec2}

As was indicated in the first section, the ground
state for the low dopings of cuprates is HTS from charged bosons. However, under 
strong magnetic fields, which were used for the measurements of the heat
conductivity  $\kappa$ and the IGS and low-$T$ MIC, the HTS is
suppressed and for PG normal state at low-$T$ we have only the
excited state -- IGBQ, which is not effected by magnetic field (see Sec. \ref{sec4} below). 
Therefore, in order to investigate the $T$
dependence of $c$ for this low doping and low-$T$, we calculate the
specific heat of IGBQ.

The thermodynamic free energy \ci{landau} of an ideal boson gas is
\be F = T \int \frac{d^2 \, p \, d^2 \, x}{(2\pi \hbar)^2}\, ln
(1-e^{-\varepsilon (p)/T}) \ . \lab{wf1} \ee Here, the temperature
$T$ is described by the energy scale. We use a relation between
the quasi-particle (QP) energy $\varepsilon$ and momentum $p$ of
IGBQ, which was derived previously \ci{Abdullaev5}: $\varepsilon
(p)=(ap+(p^2/2m)^2)^{1/2}$, where $ap$ is the square of QP energy
of plasmon, and $a=2\pi e^2 n \hbar/m$. Here $e$ and $n$ are the
charge and 2$D$ density, respectively, of real particles with mass
$m$.

One can separate the integral of Eq. \re{wf1} over the absolute
value of momentum $p$ into two parts:
$\int_{0}^{\infty}=\int_{0}^{q}+\int_{q}^{\infty}$. Here $q$ gives
the maximum of the momentum distribution function of
non-condensate real particles \ci{lifshitz} at $T=0$ under the
Bogoliubov approximation. Thus the contributions of the plasmon
(dominant for $ p < q$) and of the kinetic energy (dominant for $p >
q$) to the energy of QPs are approximately separated at the
boundary ($p=q$), where \be (aq)^{1/2}=q^2/(2m) \ . \lab{wf1a} \ee

For the first (second) integral, if we introduce the variable
$x=(ap)^{1/2}/T$ ($x=p^2/(2m T)$), we obtain
\begin{eqnarray}
F&\approx& \dsf{T S}{2\pi \hbar^2} \left[ \dsf{2T^4}{a^2}
\int\limits_0^{(aq)^{1/2}/T} d \, x
\, x^3 \, ln(1-e^{-x})\right.\nonumber\\
&+& m \left. T \int\limits_{(aq)^{1/2}/T}^{\infty} d \, x \,
ln(1-e^{-x}) \right] \, , \label{wf2}
\end{eqnarray}
where $S$ is defined to be the area of 2$D$ system.
(i) At lower temperatures, one can have $(aq)^{1/2}/T \gg 1$, and one can replace
the upper limit of first integral in Eq. \re{wf2} by infinity sign,
and the second integral becomes zero.
(ii) At high temperature, we can have $(aq)^{1/2}/T \ll 1$,
then the lower limit of
second integral can be replaced by zero and the first
integral is removed.
Following the calculational scheme of Ref. \ci{landau}, we thus find
that the specific heat per area can be expressed at low-$T$ by: \be
c_1=-\dsf{20 T^4 A}{\pi \hbar^2 a^2} \lab{wf3} \ee with the ideal gas
of the plasmons and at high-$T$ by: \be c_2=-\dsf{T m B}{\pi \hbar^2 } \lab{wf4} \ee
with the ideal gas of free QPs. Here,
$A=-(1/4)\int_0^{\infty}x^4dx/(e^x-1)=-(1/4)\Gamma(5) \zeta(5)$
and $B=-\int_0^{\infty}x dx/(e^x-1)= -\Gamma(2) \zeta(2)$, where
$\Gamma(y)$ and $\zeta(y)$ are Gamma and the Riemann Zeta
functions, respectively.

We define a temperature $T_d^c$, at which the low-$T$ dependence
of $c_1$ transforms into that of $c_2$ at high-$T$. It can
approximately be determined by setting equal the approximate equations for
$c_1$ and $c_2$. It has a expression \be T_d^c= \left( \dsf{\Gamma
(2) \zeta (2)}{5 \Gamma (5) \zeta (5)} \right)^{1/3} (ma^2)^{1/3}
\, . \lab{wf6} \ee One can show that
$(ma^2)^{1/3}=2(2/r_s^2)^{2/3}Ry$, where $r_s$ is the mean
distance between particles in Bohr $a_B$ radius
($a_B=\hbar^2/(me^2)$) and $Ry$ is Rydberg ($Ry=me^4/(2\hbar^2)$)
energy. We use the relation \ci{abdullaev1,abdullaev1a} $n_{ab}=6.747 \cdot
10^{14} \cdot t/cm^2$ between the density of charges in cuprate $1
\, cm^2$ area of $a-b$ plane and fraction $t$ of charge per $Cu$
atom. Expressing the density in $1/a_B^2$ units we find $r_s$ as
function of $t$. Therefore, $T_d^c$ is: \be T_d^c \approx 0.114
t^{2/3} Ry \, . \lab{wf7} \ee

The specific heat at an arbitrary temperature is obtained from
Eq. \re{wf1}, which has the form: \be c= \dsf{q^2}{2m} \dsf{m}{4\pi
\hbar^2} \int_0^{\infty} \dsf{d \, y}{z^2} \dsf{
y^2(1+y^3)}{sinh^2(y^{1/2}(1+y^3)^{1/2}/(2z))}\, , \lab{wf8} \ee
where $z=T/(aq)^{1/2}$ and $y=p/q$. Fig. 1 shows the $c/z$ (in
$q^2/(8\pi \hbar^2)$ units) calculated by using Eq. \re{wf8}.
Here, it is remarkable that the curve can nicely describe the
downturn behavior of the specific heat. The downturn occurs around
$z=0.5$, indicating that $T_d^c= 0.5(aq)^{1/2}$. Here, $(aq)^{1/2}
= 8(2^{1/2}/(3r_s^2))^{2/3}Ry$ and expressing $r_s$ through $t$,
$(aq)^{1/2} = 0.737\cdot t^{2/3}Ry$. By the numerical estimation
of the downturn temperature $T_d^c$, the coefficient in Eq.
\re{wf7} becomes 0.368 instead of 0.114.

\section{Heat conductivity and MIC} \label{sec3}

The downturn behavior for the heat conductivity $\kappa$ deviated
from the WFL can be explained by a crossover of freely penetrating
QPs of IGBQ into localizing individual particles collective
plasmon state. As mentioned in the first section, the $a-b$ plane 
cuprate atoms deform the intra charge structure of single bosons and 
pin them (fixed nematicity phase). However, they become 
free (free QPs) at the bosonic insulator - bosonic metal transition 
temperature and doping (see below). At the same time, despite the fact that
the charged single bosons are pinned before this transition, the 
plasmon gas is free for a penetrating. 

The plasmon gas, which becomes dominant at low-$T$, does not carry
the charge and is electrically insulating, therefore, the formation
of plasmons at low-$T$ significantly affects the behavior of
$\kappa$. One can approximately estimate the downturn temperature
$T_d^{\kappa}$ of $\kappa$  as a temperature, where WFL of
electrically conducting free QPs  with electrical conductivity
$\sigma_2$  transits into WFL of plasmons (with the electrical
conductivity $\sigma_1$). We apply the WFL in the region of this
nearly insulating gas (mainly consisted of plasmons) to calculate
$\sigma_1$ from the $\kappa$ and specific heat. Although the WFL
is not conventionally applicable for insulators, we will
demonstrate in this section that in the case of cuprates
$\sigma_1$ for plasmons equals $\sigma_2$ of conductor with
infinitely small numerical prefactor. Thus for plasmon gas as
conductor with very small electrical conductivity the WFL is
formally valid. On the other hand, this "bad" conductor satisfies
the formal definition of insulator given in the classical
electrodynamics as conductor having its conductivity close to zero.
While the Lorentz ratio in our paper calculates only for free QPs
of IGBQ, i.e., for conductor above the downturn $T_d^{\kappa}$
temperature. We cannot, strictly saying, apply the WFL for
insulating plasmon gas to calculate the Lorentz ratio. This our
result for Lorentz ratio, Fig. 2 is in correspondence with
experiment of Proust $et$ $al.$ \ci{proust}.

According to the Drude model, which we will apply for the description
of heat conductivity $\kappa$, only two quantities (except the
specific heat $c$) contribute in $\kappa$ (see below): mean
velocities and free penetration lengths of QPs. For low-$T$
regime, at which the experiments on $\kappa$ are measured, one can
ignore the $T$ dependence of these two (if the mean velocity
is determined by concentration of charges, then the free
penetration length is determined for low-$T$ by scattering of
charges on impurities). Therefore, the qualitative $T$ dependence
of $\kappa$ (together with downturn $T_d^{\kappa}$) is determined
only by the specific heat $c$.

Following the Drude model, we have $\kappa_1 (T)=(1/2)c_1(T)v_1
l_1$ for gas of plasmons and $\kappa_2 (T)=(1/2) c_2(T)v_2 l_2$
for gas of free QPs with corresponding mean velocities and free
penetration lengths $v_{1,2}$, and $l_{1,2}$, respectively. To obtain
the value of $T_d^{\kappa}$, we express the specific heat Eq.
\re{wf8} \be c_1= \dsf{q^2}{2m} \dsf{m}{\pi \hbar^2} z^4 10 \Gamma
(5) \zeta (5)\, \lab{wf8p1} \ee for $z\ll 1$ and \be c_2=
\dsf{q^2}{2m} \dsf{m}{\pi \hbar^2} z \Gamma (2) \zeta (2)\,
\lab{wf8p2} \ee for $z\gg 1$. Substituting $z=T/(aq)^{1/2}$ and
using Eq. \re{wf1a} it is found that Eq. \re{wf8p1} reduces to Eq.
\re{wf3} and Eq. \re{wf8p2} to Eq. \re{wf4}.

Therefore, WFL is expressed as: \be \dsf{\kappa_1}{\sigma_1 T} =
\dsf{k_B m v_1 l_1}{2\pi \hbar^2 \sigma_1}  z^3 10 \Gamma (5)
\zeta (5)\, \lab{wf8p3} \ee for gas of plasmons  and \be
\dsf{\kappa_2}{\sigma_2 T} = \dsf{k_B m v_2 l_2}{2\pi \hbar^2
\sigma_2} \Gamma (2) \zeta (2)\, \lab{wf8p4} \ee for gas of free
QPs.

At the downturn temperature $T_d^{\kappa}$, it should be $\kappa_1
(T_d^{\kappa})/(\sigma_1 T_d^{\kappa})=\kappa_2
(T_d^{\kappa})/(\sigma_2 T_d^{\kappa})$. However, due to the
non-equality $v_1 l_1/\sigma_1 \not= v_2 l_2/\sigma_2 $ we cannot
express Eq. \re{wf8p3} and  Eq. \re{wf8p4} as single function of
WFL, which transforms from low-$T$ limit into high-$T$ limit with
the increase of $T$. On the other hand, if we write $v_1
l_1/\sigma_1 = K_{\sigma_1,\sigma_2} v_2 l_2/\sigma_2$, where
$K_{\sigma_1,\sigma_2}$ is numerical factor, and if we introduce
the definition $v l/\sigma = v_2 l_2/\sigma_2$ for gas of free
QPs, the single WFL for new parameter can be expressed by
$z=K_{\sigma_1,\sigma_2}^{1/3}T/(aq)^{1/2}$. This can describe the
WFL of the free QPs gas at all $T$. In Fig. 1, we plot the
$\kappa/(\sigma T)$, (expressed in $k_B m v l/(2\pi \hbar^2
\sigma)$ units) as function of $z$, which is valid at low-$T$,
where $v$, $l$ and $\sigma$ are $T$ independent.
\begin{figure}
\begin{center}
\includegraphics[angle=0,width=8.5cm,scale=1.0]{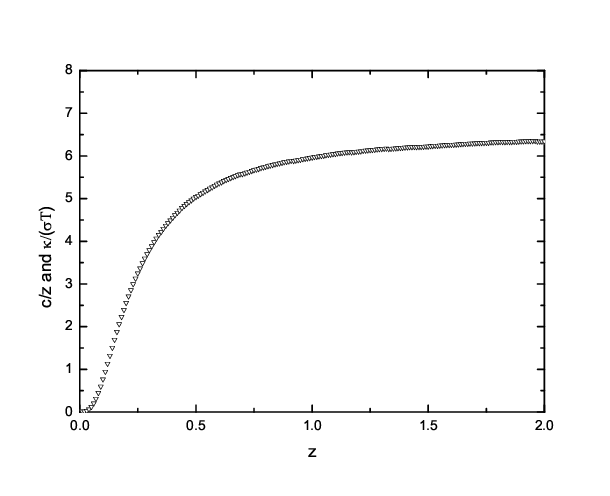}
\end{center}
\caption{ The specific heat $c/z$ (in $q^2/(8\pi \hbar^2)$ units)
and $\kappa/(\sigma T)$ (in $k_B m v l/(2\pi \hbar^2 \sigma)$
units) as function of $z$. The definition of $z$: $z=T/(aq)^{1/2}$
for $c/z$ and $z=K^{1/3} T/(aq)^{1/2}$ for $\kappa/(\sigma T)$,
where $K\sim 10^{12}$ (see text). } \lab{fig1}
\end{figure}

We estimate the electrical conductivity $\sigma_1$ for a case of
$t\sim 0.1$. From the expression for $(aq)^{1/2}$, one derives
$(aq)^{1/2}/k_B \sim 10^4 \, K$ (in $K$ -- Kelvin temperature
units). Thus, the downturn of the specific heat takes place at
$T_d^c \sim 10^4 \, K $. The observed downturn-$T$ of the heat
conductivity \ci{hill,proust} is $T_{d,exp}^{\kappa}/k_B \sim 0.1
\, K $. By using the downturn coordinate $z$ = 0.5 from Fig. 1 for
$\kappa/(\sigma T)$ and by substituting in this $z$ values of
$(aq)^{1/2}/k_B$ and $T_{d,exp}^{\kappa}/k_B$, we obtain
$K_{\sigma_1,\sigma_2}\sim 10^{12}$. This means that $\sigma_1 v_2
l_2/(\sigma_2 v_1 l_1)\sim 10^{-12}$. We compare $v_1$ with $v_2$
and $l_1$ with $l_2$. Free QPs crossover into plasmons at momentum
$q$ determined by Eq. \re{wf1a}. However, this expression for $q$
is similar to the expression for the critical momentum for the Landau
damping of electrons \ci{ichimaru}, which decay plasmons being generated
or absorbed. Thus, we can apply the Landau damping
approach in our case. However, except an energy conservation law
-- Eq. \re{wf1a}, this approach requires an equality of the plasmon phase
velocity and electron velocity. Therefore, for an
estimate, one can assume at $T_d^{\kappa}$  that $v_1 \sim v_2$
and $l_1 \sim l_2$ (here we take that the plasmon phase
velocity is roughly equal to the plasmon group velocity and the lifetime
of plasmon and electron against to decay is the same), i.e., mean
velocities and free penetration lengths for plasmons and free QPs
have the same order of magnitude. Hence, we obtain $\sigma_1
/\sigma_2 \sim 10^{-12}$. This result supports the above supposed
assumption of the IGS, in which the insulating plasmon gas formed
from the charge conducting free QP gas transits at low-$T$.

This analysis allow us to suggest that the downturn of heat
conductivity may be a result of the MIC at low-$T$. In addition to
coincidence of experimental parameters for the doping and of
magnetic field strength at the measurements of $\kappa$ and the
IGS, and low-$T$ MIC for the same copper-oxide, as discussed above,
the MIC boundary described by Boebinger {\it et} {\it al.} (see Ref.~\ci{Boebinger})
(at higher $T$) has qualitatively the same doping dependence as
the PG boundary. In Ref.~\ci{abdullaev1}, we limited the existence
region of anyon-related Coulomb interacting bosons below the PG
boundary because close to PG boundary bosons were transformed to
fermions. This observation of Boebinger {\it et} {\it al.} supports the
frame of our approach on the nature of PG phase as a mixture of
single particle bosons and normal fermions. In our description the
region below Boebinger {\it et} {\it al.} MIC boundary is dominated by plasmons,
while close and right above of this boundary by free QPs
of IGBQ. 

This scenario for MIC corresponds to the bosons of the Coulomb 
single boson and single fermion two liquid model and fermions, whose origin is 
these bosons. At an increase of doping up to critical value or temperature up 
to PG boundary temperature, this boson system undegoes bosonic insulator - bosonic 
metal - fermionic metal transitions. While there also exists a small part of
the model fermion component, which is not related to single bosons and at the same 
variation of doping or temperature undegoes insulator - metal crossover~\ci{Abdullaev1,Abdullaev7}
and does not contribute to the MIC. 

We determine the Boebinger {\it et} {\it al.} MIC boundary for temperature and doping
in the doping-temperature phase diagram. This boundary defines temperature as 
function of doping, at which bosonic insulator - bosonic metal transition occurs.
The analytic form for the transition temperature is expressed by Eq.~\re{wf7}  
with coefficient 0.368 instead of 0.114 for the downturn temperature $T_d^c$ of the specific
heat. In the next paper, part II, we will demonstrate that the low-$T$ dependence of a
resistivity is determined by the specific heat.
However, in Eq.~\re{wf7}, we used the expression for the entire density of charges
$n_{ab}$. While one needs to separate in it the contributions from single boson and single
fermion parts. We write the formula $n_{ab}=n_{ab}(1-t/t_c)+n_{ab}t/t_c$, in which 
the first term describes the approximate single boson density and the second one
the approximate single fermion density. Our interest is in the single bosons therefore,
the expression for the bosonic insulator - bosonic metal transition temperature 
is: \be T_{bIM} \approx 0.368 
[t(1-t/t_c)]^{2/3} Ry \, . \lab{wf7a} \ee Eq.~\re{wf7a} describes the temperature and doping
boundary for the MIC of resistivity, found in the experiment of Boebinger {\it et} {\it al.}~\ci{Boebinger}.

We now calculate the Lorentz ratio for free QPs of IGBQ. The WFL,
$ \kappa/(\sigma T)=L_0$, of heat transport for QPs of LFLT  can
be expressed by the Lorentz ratio $L_0$ and Sommerfeld's value
$L_0=(\pi^2/3) (k_B/e)^2$. We note that this value of $L_0$
corresponds to the three dimensional ($3D$) case. A simple
calculation shows that it can be also applied for $2D$ case. For
the IGBQ, it can be assumed that the mean kinetic energy of QPs is
$mv^2/2= (aq)^{1/2}$ (Eq. \re{wf1a}, because, as was pointed out
above, at $T=0$ the most part of non-condensate particles has this
energy). Then $\kappa/(\sigma T)= mv^2c/(2ne^2T)$, with $c$
determined from Eq. \re{wf4} for free QPs of IGBQ, reduces to a
form $\kappa/(\sigma T)= 3.106\cdot L_0/t^{1/3}$. This expression
corresponds to WFL of bosons, when there is no mixture of bosons
with Fermi QPs of LFLT. However, at the increasing of
concentrations of holes or electrons $t$ to the direction of the 
critial doping $t_c$, fermions appear in the PG region
\ci{abdullaev1} and this mixture occurs. Therefore, at
$T\rightarrow 0$ we can phenomenologically write the expression
$\kappa/(\sigma T)=L$ with: \be L = L_0 \left[\dsf{3.106}{t^{1/3}}
\left(1-\dsf{t}{t_c} \right)+\dsf{t}{t_c}\right] \, , \lab{wf9}
\ee  which takes into account the transformation of $L$ from one
of free QPs of IGBQ to one of LFLT QPs, when the concentration $t$
tends to (but below) critical concentration $t_c\approx 0.19$
\ci{tallon}. In Fig. 2 we compare our estimated $L/L_0$ with the
experimental data taken from Fig. 7 (a) of \ci{proust} for several
cuprate compounds. It is remarkable that the calculated curve is
in an good agreement with the observed values of $L/L_0$.
\begin{figure}
\begin{center}
\includegraphics[angle=0,width=8.5cm,scale=1.0]{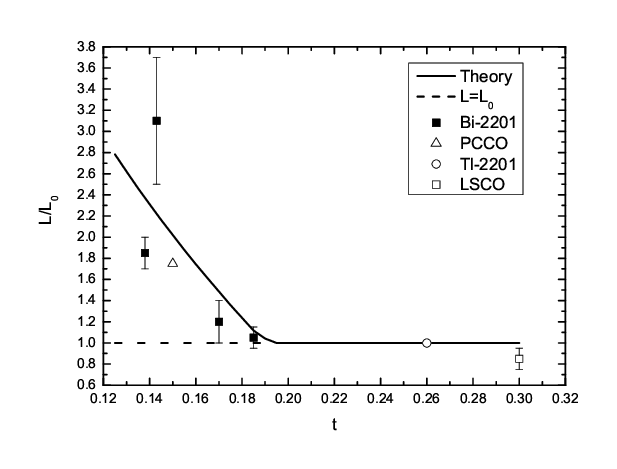}
\end{center}
\caption{ The Lorentz ratio $L/L_0$ (Eq.~\protect\re{wf9})
vs. $t$ (values for $t\ge 1$ are added artificially). Observed
dots are from Ref. \protect\ci{proust}.} \lab{fig2}
\end{figure}

\section{Specific heat and entropy}
\label{sec4}

The qualitative features \ci{loram1,loram2,tallon} of the
observed cuprate electronic specific heat and entropy are as follows. The
increment coefficient $\gamma=c/T$ of the specific heat in the
HTS-to-normal-state phase transition (PT) point has weakly
apparent and washed out peak in the lightly underdoped side. The
peak of $\gamma$ and its form become higher and sharper,
respectively, when $t$ increases. Peak is maximal (with the
sharpest form) at $t_c$ and then does not almost change the form
for $t>t_c$. There is no influence on the normal state $c$ by the
external magnetic field \ci{luo}. At last, $\gamma$ is independent
from doping for $t>t_c $ \ci{loram1}.

In general, the weakly apparent peak of $\gamma$ is attributed
\ci{landau} to the first order PT, which is close to the second
order one. Sharpening of $\gamma$ at $t_c$ might characterize the
transition of PT into the second order type, where conventional
superconductivity scenario with Cooper pairs becomes effective.
The hypothesis in the accordance with this description was pointed
out in \ci{abdullaev1}.

The non-dependence of the normal state $c$ on the magnetic field
might be a result that QP energy $\varepsilon (p)$ of IGBQ, by
analogy with Cooper pair energy $\varepsilon_{Cp} (p)$, is
independent of magnetic field. Abrikosov \ci{abrikosov} has shown
that $\varepsilon_{Cp} (p)$ being the result of canonical
Bogoliubov transformation is function of scalar quantities $u_p$
and $v_p$. However, the latter ones, as scalars, in the magnetic
field with gauge $\vec \nabla \cdot \vec A(\vec r)=0$ can be
function of only zero scalar product $\vec k\cdot \vec A_{\vec
k}=0$, where $\vec k$ and $\vec A_{\vec k}$ are Fourier transforms
of vector-coordinate $\vec r$ and vector-potential $\vec A (\vec
r)$, respectively. The independent from the magnetic field PG normal
state was also observed in the resistivity measurement of the MIC
 \ci{daou}, thus sustaining the possible role of $\varepsilon
(p)$ of IGBQ.

For the specific heat $c_F$ of $2D$ gas of LFLT QPs one obtains
the same linear $T$ dependence as in Eq. \re{wf4}, but with ratio
$c_F/c_2=2$ (due to two directions of spins for fermions instead
of bosons). The non-$t$-dependence observation of $\gamma$ for
$t>t_c$ might be a result of the non-concentration dependence for Eq.
\re{wf4}.

We obtain the expression of the electronic normal state entropy ${\cal S}$
through the calculation of integral ${\cal S}=\int_{0}^{T} (c/T_1)
d \, T_1 \,$ for PG, $T\le T^*$, region. Here $T^*$ is temperature
of PG boundary. Typical experimental scale of $T^*$ is
$T^*/k_B\sim 10^2 \, K$, while $(aq)^{1/2}/k_B \sim 10^4 \, K$,
therefore, in Eq. \re{wf8} one can assume $z\ll 1$ and use Eq.
\re{wf3} as approximate expression of the specific heat of IGBQ.
For arbitrary $T$ (in the interval of $ T\le T^*$), we write the
phenomenological expression: \be
\dsf{c}{T}=\dsf{c_1}{T}\left(1-\dsf{5T}{4T^*}\right)+\dsf{2c_F}{T}
\dsf{T}{T^*} \, . \lab{wf10} \ee The factor 5/4 in front of first
$T/T^*$ term was introduced for convenience purpose. The
integration over $T_1$ gives: \be {\cal
S}=\dsf{c_1}{4}\left(1-\dsf{T}{T^*}\right)+c_F \dsf{T}{T^*} \, .
\lab{wf11} \ee

We compare the $T$ dependence of our ${\cal S}$ with experimental
one of \ci{loram1}. It is convenient to express ${\cal S}$ in
$mJ/(mol K)$ units and $T$  in $K$ units (we use the approximate
PG boundary $T^*=900-4736.8421\cdot t$ taken from Fig. 11 of Ref.
\ci{tallon} and at calculating of ${\cal S}$ we assume that value
of charge is $t$). In this case, the increment coefficient
$\gamma$ from Fig. 4 of \ci{loram1}, being multiplied by $T$,
gives ${\cal S}$ as function of $T$, analogous to Fig. 6 of
\ci{luo}. However, if Fig. 6 describes only the metallic $t$ of
holes, we obtain the data and for the insulating $t$. Comparing in
Fig. 3 the plot of ${\cal S}$ obtained from Eq.~\re{wf11} with one
from Fig. 6 of \ci{luo} we see (i) the general nonlinear, $ \sim
T^i$ with $i>1$, behavior of all curves for $t<t_c$, (ii) all
curves for $T>T^*$ have a linear behavior with the same slope,
$\gamma \approx 1.46$,  as for 2$D$ fermion gas, (iii) in contrast
to experimental result our curves for $T>T^*$ are not parallel.
However, the parallelism of the observed curves indicated  in the last
point is inconsistent with the clear tendency of $\gamma$ to
approach the fixed value as $T$ goes to infinity (more obviously
it is seen in \ci{loram2} for $La_{2-x} Sr_x CuO_4$ compound,
although, in this paper $\gamma \approx 1.0$ was obtained).
Finally, the alternative $T/T^*$ dependencies are considered in
\ci{kim} for $T$ just after $T_c$ of HTS and near the optimal
doping, and in \ci{warma} close to $T^*$.
\begin{figure}
\begin{center}
\includegraphics[angle=0,width=8.5cm,scale=1.0]{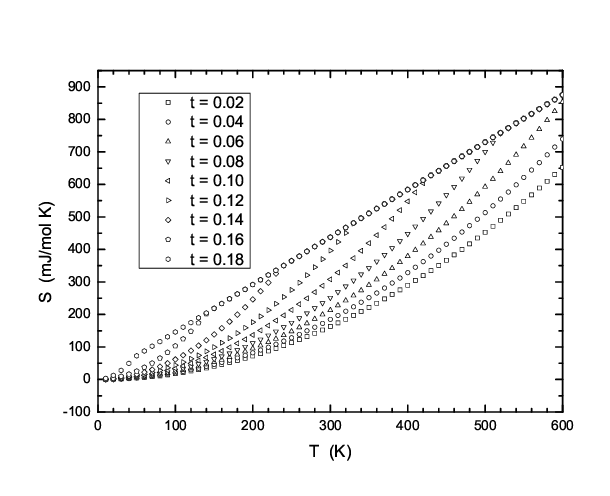}
\end{center}
\caption{ The entropy ${\cal S}$ (Eq.~\protect\re{wf11}) vs. $T$
at various $t$ (values of ${\cal S}$ behind the crossing of linear
and nonlinear parts of ${\cal S}$ are added artificially).}
\lab{fig3}
\end{figure}
However, irrespective the quality of $t/t_c$ and $T/T^*$ laws in
Eqs. \re{wf9} and \re{wf11} one can say that at small values for these
parameters the single particle boson contribution into $\kappa$
and ${\cal S}$ properly describes the experiment.

\section{Summary and conclusion}
\label{sec5} We have carried out an attempt to describe in the
unified anyon-related boson approach the PG phase electronic
low-$T$ heat conductivity $\kappa$, entropy ${\cal S}$ and the IGS,
and low-$T$ MIC of cuprates. We have argued that the observed $\kappa
\sim T^{3.6}$ and $\kappa \sim T$ is a result of $c \sim T^4$ and
$c \sim T$ dependencies, respectively, while ${\cal S} \sim T^i$
with $i>1$ originates from $c \sim T^4$ of the specific heat $c$
for IGBQ of 2$D$ Coulomb-interacting boson gas. Providing by the
qualitative and quantitative arguments, we have attributed the
downturn behavior of $\kappa$ to the MIC and transition into the
IGS. Assuming that the PG phase consisted from the mixture of IGBQ
and 2$D$ LFLT QP gas, we have obtained the Lorentz ratio values of
WFL, which were close to experimental ones. We have clarified the
origin of the decoupling of charge carriers with phonons as
transition of free QPs of IGBQ into plasmons.

We have shown that the total MIC originates from bosons of the Coulomb 
single boson and single fermion two liquid model and fermions, whose origin is 
these bosons. At an increase of doping up to the critical value (Eq.~\re{wf9}) or temperature up 
to PG boundary temperature (Eq.~\re{wf11}), this boson system undegoes bosonic insulator - bosonic 
metal - fermionic metal transitions. It is interesting that this MIC behaviour of the  
doping and temperature resembles the right-angled triangle rule, in which variables 
vary along horizontal and vertical legs, respectively. We note that there also exists a 
small part of the model fermion component, which is not related to single bosons and 
at the same variation of doping or temperature undegoes insulator - metal crossover
and does not contribute to the MIC (see Ref.~\ci{Abdullaev1}). We have succeeded in
describing of the Boebinger {\it et} {\it al.} experimental temperature and doping MIC boundary 
of resistivity.

The good agreement of the calculated within Coulomb two liquid model results for 
heat conductivity downturn, anomalous Lorentz ratio, and nonlinear entropy with
experimental ones may indicate on the correctness of this model's concepts. In
paper II of this series, we will try to understand within our model the physics of other MIC 
phenomena: resistivity temperature upturn, insulating ground state, 
nematicity- and stripe-phases, and Fermi pockets. The next important issue obtained in
paper II  will be a demonstration that the recently observed
in the hidden magnetic order and STM experiments spin and charge fluctuations 
are the intra PG and HTS pair fluctuations.

Some results of the presented paper, without derivation and detail discussion, have 
partially been published in Ref.~\ci{Abdullaev7}. 

\section{Acknowledgement}
\label{sec6}

Authors B. Abdullaev and C. -H. Park acknowledge the support of the research 
by the National Research Foundation (NRF) Grant (NRF-2013R1A1A2065742) of the 
Basic Science Research Program of Korea.

\end{document}